\documentclass[sigconf]{acmart}

\copyrightyear{2024}
\acmYear{2024}
\setcopyright{acmlicensed}\acmConference[WWW '24]{Proceedings of the ACM Web Conference 2024}{May 13--17, 2024}{Singapore, Singapore}
\acmBooktitle{Proceedings of the ACM Web Conference 2024 (WWW '24), May 13--17, 2024, Singapore, Singapore}
\acmDOI{10.1145/3589334.3645347}
\acmISBN{979-8-4007-0171-9/24/05}
 
\usepackage{amsmath}               
  {
      \theoremstyle{plain}

  }
\usepackage{enumitem}
\AtBeginDocument{%
  \providecommand\BibTeX{{%
    \normalfont B\kern-0.5em{\scshape i\kern-0.25em b}\kern-0.8em\TeX}}}

\usepackage{tikz}
\usepackage{subcaption}
\usepackage{pgfplots}
\usetikzlibrary{arrows,positioning,automata,calc,shapes}
\pgfplotsset{compat=newest, scaled z ticks=false} 
\pgfplotsset{plot coordinates/math parser=false}
\newlength\figureheight 
\newlength\figurewidth
\usepackage{tikz}
\usepackage{caption}
\usepackage{filecontents}
\usepackage{url}
\usepackage{amsmath}
\usepackage{graphicx}
\usepackage{url}
\usepackage{setspace}
\usepackage{hyperref}

\usepackage{amsfonts,amssymb}
\usepackage{color, soul}
\usepackage{mathrsfs}
\usepackage{cleveref}
\usepackage{multirow}
\usepackage{wrapfig}
\usepackage{amsthm}
\usepackage{algorithm}
\usepackage{algorithmic}
\usepackage{bm}
\usepackage{mdframed}

\begin{document}

\setlength{\fboxsep}{1.5pt}

\title{Collaborative Large Language Model for Recommender Systems}


\author{Yaochen Zhu}
\authornote{Work done when Yaochen Zhu was an applied research intern at LinkedIn.} 
\affiliation{%
\institution{University of Virginia}
\city{}
\state{}
\country{}
}
\email{uqp4qh@virginia.edu}

\author{Liang Wu}
\affiliation{%
\institution{LinkedIn Inc.}
\city{}
\state{}
\country{}
}
\email{liawu@linkedin.com}

\author{Qi Guo}
\affiliation{%
\institution{LinkedIn Inc.}
\city{}
\state{}
\country{}
}
\email{qguo@linkedin.com}

\author{Liangjie Hong}
\affiliation{%
\institution{LinkedIn Inc.}
\city{}
\state{}
\country{}
}
\email{liahong@linkedin.com}

\author{Jundong Li}
\affiliation{%
\institution{University of Virginia}
\city{}
\state{}
\country{}
}
\email{jundong@virginia.edu}

\begin{abstract}

Recently, there has been growing interest in developing the next-generation recommender systems (RSs) based on pretrained large language models (LLMs). However, the semantic gap between natural language and recommendation tasks is still not well addressed, leading to multiple issues such as spuriously correlated user/item descriptors, ineffective language modeling on user/item data, inefficient recommendations via auto-regression, etc. In this paper, we propose \textbf{CLLM4Rec}, the first generative RS that tightly integrates the LLM paradigm and ID paradigm of RSs, aiming to address the above challenges simultaneously. We first extend the vocabulary of pretrained LLMs with user/item ID tokens to faithfully model user/item collaborative and content semantics. Accordingly, a novel \textit{soft+hard prompting} strategy is proposed to effectively learn user/item collaborative/content token embeddings via language modeling on RS-specific corpora, where each document is split into a prompt consisting of heterogeneous \textit{soft} (user/item) tokens and \textit{hard} (vocab) tokens and a main text consisting of homogeneous item tokens or vocab tokens to facilitate stable and effective language modeling. In addition, a novel mutual regularization strategy is introduced to encourage CLLM4Rec to capture recommendation-related information from noisy user/item content. Finally, we propose a novel recommendation-oriented finetuning strategy for CLLM4Rec, where an item prediction head with multinomial likelihood is added to the pretrained CLLM4Rec backbone to predict hold-out items based on soft+hard prompts established from \textit{masked} user-item interaction history, where recommendations of multiple items can be generated efficiently without hallucination\footnote{Codes are released at \url{https://github.com/yaochenzhu/llm4rec}.}. 

\end{abstract}

\keywords{Recommender systems; large language models (LLM)}

\ccsdesc[500]{Information systems~Recommender systems}

\maketitle

\vspace{-3mm}

\section{Introduction}

With content growing exponentially on the Web, recommender systems (RS) have become essential components for online service platforms  \citep{jannach2010recommender}. Nevertheless, RS has long been dominated by the ID-based paradigm, where users/items are represented by unique, continuous ID embeddings denoting their semantic similarity \cite{yuan2023go}. Exemplar ID-based RSs include matrix factorization-based methods (such as PMF \citep{mnih2007probabilistic}) and two-tower models \citep{wu2018starspace}, where user/item ID embeddings are either randomly initialized and learned from their historical interactions (i.e., collaborative filtering \citep{koren2021advances}), or established based on user/item features (i.e., content-based methods \cite{lops2011content,Yaochen}).

\begin{figure}[t]
\centering
\includegraphics[width=0.78\linewidth]{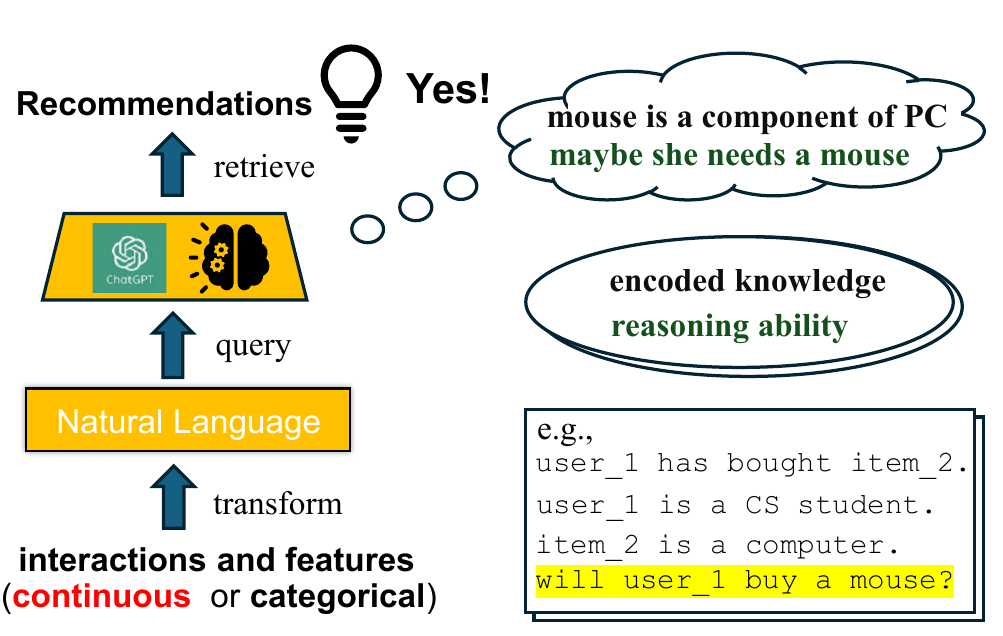}
\vspace{-2mm}
\caption{Prospectives of developing the next generation of recommender systems based on pretrained LLMs.}
\label{fig:teaser}
\vspace{-5mm}
\end{figure}

Recently, large language models (LLM) have become a heated topic for both academia and industry \cite{zhao2023survey}. Large transformer networks pretrained on large-scale corpora, such as GPT \cite{radford2018improving}, T5 \cite{raffel2020exploring}, and LLaMA \cite{touvron2023llama}, have demonstrated \textbf{emergent ability} \cite{wei2022emergent}, showcasing unprecedented understandings of knowledge and patterns in natural language \cite{zhao2023survey,wang2023knowledge}. Consequently, it is promising to develop the next generation of RS based on pretrained LLMs \cite{fan2023recommender}, fully utilizing their encoded knowledge, logical reasoning ability, and generative AI power to understand and reason with user/item semantics and make more accurate recommendations accordingly, especially when users and items are associated with large amounts of textual features, such as biographies, descriptions, content, reviews, and explanations, in modern online service platforms \cite{mcauley2016addressing,zhu2022variational}. 

Several preliminary studies have been conducted to explore the adaptation of LLMs for RSs \cite{cui2022m6,geng2022recommendation,qu2023language,li2023personalized}. Typically, these methods can be summarized into two steps: \textbf{\textit{(i)}} First, instead of representing users/items with continuous ID embeddings, relevant information necessary for reasoning with user interests and generating recommendations, e.g., interacted items, user/item features, and candidate items, is converted into a discrete \textit{natural language-based prompt}. \textbf{\textit{(ii})} Then, the prompt is used to query the LLM, where information relevant to recommendations is retrieved from the \textit{textual output} of the LLM to generate recommendations (see Fig. \ref{fig:teaser} for an intuitive example).
The above procedure can be performed in a zero-shot manner \cite{gao2023chat,hou2023large,zhang2023recommendation,he2023large}, where the recommendation decisions are obtained directly from the pretrained LLM (e.g., we input all relevant information regarding a user and an item into the chatbox of ChatGPT and ask if the user will interact with the item), or if the groundtruths are available, the pretrained LLMs can also be finetuned on both interaction and feature data, such that RS-specific knowledge can be incorporated for more accurate recommendations \cite{geng2022recommendation,yang2023palr,ji2023genrec,chu2023leveraging}. 

Although impressive progress has been achieved, fundamental dichotomies between NLP and recommendation still remain to be addressed. One main challenge is the gap between natural language and user/item semantics. Generally, there are two strategies to represent users/items in an LLM-based RS. Pseudo-ID-based methods use an ID-like word  (e.g., "user\_$i$" or "item\_$j$") to represent the $i$-th user or $j$-th item \cite{geng2022recommendation}. However, when tokenized, the ID word may be broken down into atomic tokens, e.g., "user\_4332" into ["user", "\_", "43", "32"], where spurious correlations can be introduced for irrelevant users/items (e.g., "user\_4332" with "user\_43" and "user\_32"). In contrast, description-based methods use semantically meaningful tokens to index users/items, such as item titles \cite{cui2022m6,hou2023large} or a small amount of newly-introduced tokens assigned to different users/items based on content similarity \cite{hua2023index}. However, description-based methods introduce a strong inductive bias on user-item semantic similarity, which may not faithfully capture the true semantics. Introducing true user/item ID tokens, unfortunately, is generally considered infeasible for LLMs, as directly conducting language modeling (LM) on sequences with heterogeneous tokens can be ineffective and unstable, especially when the vocabulary of most LLMs can be diluted (e.g., $\sim$ 50k for GPT, and $\sim$ 30k for T5) by a large number of randomly initialized user/item embeddings.

Even if user/item ID token embeddings can be effectively learned via LM, more challenges exist that hinder effective and efficient recommendations with LLMs. First, since the interaction order usually does not matter for direct recommendations while human language naturally has an order, spurious temporal correlation can be introduced for items placed in different positions when transforming user historical interactions into a textual sentence. In addition, for content modeling, since pretrained LLMs are not recommendation-oriented, they can easily capture noise in user/item textual features irrelevant to the recommendation purpose. Furthermore, since LLMs generate the next token in an autoregressive manner, making multiple recommendations via LLM-based RSs can be inefficient compared with ID-based methods. Finally, for both pseudo-ID-based and description-based indexing methods, item candidates usually need to be explicitly provided in the prompt to avoid hallucination \cite{geng2022recommendation}. These issues hinder the practical applications of LLM-based RSs where candidate pools are large and low latency matters.

To address the above challenges, we present \textbf{CLLM4Rec}, the first generative RS that tightly combines the ID paradigm of RS with the LLM-based paradigm. We first extend the vocabulary of pretrained LLMs with user/item ID tokens to faithfully model the user/item collaborative/content semantics, where the token embeddings are learned in two stages. The \textit{pretraining stage} consists of mutually regularized collaborative or content LLMs that learn user/item token embeddings via language modeling on RS-specific corpora established from user/item interactions and textual features. Specifically, a novel "soft+hard" prompting strategy is proposed for effective language modeling on documents with heterogeneous tokens, where each document is decomposed into a prompt consisting of \textit{soft} \cite{lester2021power}  (user/item) and \textit{hard}  (vocab) tokens and a main text consisting of homogeneous item tokens (for collaborative modeling) or vocab tokens (for content modeling), respectively. Through this strategy, the prediction heads for the two LLMs can focus exclusively on collaborative and content information, such that the stability and effectiveness of language modeling can be substantially enhanced. In addition, a stochastic item reordering strategy is proposed for the collaborative LLM to ignore the order of item tokens without negative influence on the vocab tokens. Finally, we propose a novel recommendation-oriented \textit{finetuning strategy} for CLLM4Rec, where an item prediction head with multinomial likelihood is added to the pretrained collaborative LLM backbone to predict hold-out items based on soft+hard prompts established from masked user interaction history, where recommendations of multiple items can be efficiently generated without hallucination. The contribution of this paper can be concretely summarized as:
\begin{itemize}[leftmargin=0.5cm]
    \item We present CLLM4Rec, the first generative RS that tightly couples the ID paradigm and LLM paradigm, where user/item ID token embeddings aligned to the LLM vocab space are introduced to well capture the intrinsic user interests and item properties.
    \item A novel soft+hard prompting strategy is proposed to effectively pretrain CLLM4Rec on heterogeneous tokens describing historical interactions and user/item features in a mutually regularized manner, where collaborative and content information can be effectively learned by the user/item token embeddings.
    \item A recommendation-oriented finetuning strategy is proposed that predicts hold-out items based on soft+hard prompts established from masked interactions via an item prediction head with multinomial likelihood, where recommendations for multiple items can be generated efficiently without hallucination.
\end{itemize}


\section{Related Work}

\subsection{Large Language Model (LLM) Basics}

Large transformer networks \cite{vaswani2017attention} trained on large corpora, i.e., large language models (LLMs), have demonstrated unprecedented understandings of natural language and logical reasoning ability \cite{zhao2023survey}. According to the part of transformer utilized for language modeling, existing LLMs can be categorized into three classes: \textbf{\textit{(i)}} encoder-only LLMs, such as BERT \cite{kenton2019bert}, \textbf{\textit{(ii)}} encoder-decoder-based LLMs, such as T5~\cite{raffel2020exploring}, and \textbf{\textit{(iii)}} decoder-only LLMs, such as GPT, LLaMA \cite{radford2018improving, touvron2023llama}. We focus on LLMs with decoders due to their superior generative abilities compared with the encoder-only models \cite{min2023recent}. The training of LLMs is mainly based on two stages. In the pretraining stage, LLMs are trained on large corpora via language modeling (i.e., next/masked token prediction), where knowledge can be effectively encoded in the transformer network weights facilitated by the stacked self-attention modules. Then, during the finetuning stage, exemplar prompt-output pairs or human feedback on multiple generated answers are provided to the LLMs such that they can conduct logical reasoning and generate answers according to prompt based on the encoded knowledge from the pretrained stage.

\subsection{LLM in Recommender Systems}
\label{sec:llm4rec}

Recently, LLM-based RSs have shown potential to address the long-standing issues of ID-based RSs, such as shallow understanding of user/item textual features \cite{liu2023pre}, poor generalization \cite{lin2023can}, etc. Hou et al. \cite{hou2023large} demonstrated that existing LLMs can be viewed as zero-shot rankers, which can sort the relevance of movies based on user historical interactions and movie descriptions. Recently, more efforts have been devoted to the finetuning of LLMs to obtain recommendation-oriented models. An exemplar work is P5 \cite{geng2022recommendation}, which finetunes T5 on corpora established from both interactions and user/item features, where items are presented by pseudo-IDs. Afterward, M6 \cite{cui2022m6} was proposed to combine text infilling and auto-regression tasks in the pretraining stage, where pseudo IDs are replaced by textual descriptions. Recently, TALLRec \cite{bao2023tallrec} was proposed where items are represented by both pseudo-ID and textual descriptions. However, pseudo-ID-based item representations can introduce spurious correlations between irrelevant items. To address this issue, Hua et al. \cite{hua2023index} proposed to introduce a small number of new tokens to describe the items, which are determined by their content and collaborative similarity. However, indexing items with shared tokens can still introduce bias. In addition, candidate items need to be explicitly provided in the prompt, and recommendations are generated via inefficient auto-regression. In summary, the dichotomy between NLP and RS is still not well-addressed. 

\section{Methodology}

\subsection{Problem Formulation}

In this paper, we focus on recommendations with implicit feedback \cite{hu2008collaborative}. Consider a system of $I$ users and $J$ items. We use a binary rating vector $\mathbf{r}_{i} \in \{0, 1\}^{J}$ to denote whether user $i$ has interacted with the $J$ items. In addition, we use $\mathbf{x}^{u}_{i}$, $\mathbf{x}^{v}_{j}$ to denote the textual features associated with user $i$ and item $j$, such as user biography and item content, etc. $\mathbf{x}^{uv}_{ij}$ denotes the textual features associated with both user $i$ and item $j$, such as user $i$'s review for item $j$, etc. Hereafter, we take a sequential view of $\mathbf{x}^{\{u,v, uv\}}_{\{i,j,ij\}}$, where $\mathbf{x}^{\{u,v, uv\}}_{\{i,j,ij\},k}$ is a size $N$ one-hot vector denoting the $k$-th token in the textual sequence. In addition, we have a pretrained large language model (LLM), of which we take a probabilistic view and denote it as $p_{llm}(\mathbf{x}_{k+1}|\mathbf{x}_{1:k})$. $p_{llm}$ transforms $\mathbf{x}_{1:k}$ into a latent sequence $\mathbf{h}^{(L)}_{1:k} \in \mathbb{R}^{k \times K_{h}}$ via $L$ stacked self-attention modules $llm(\mathbf{x}_{1:k})$ and maps $\mathbf{h}^{(L)}_{k}$ to the probability space of the next token $\mathbf{x}_{k+1}$. Since the LLM is pretrained on large corpora and finetuned on exemplar prompt-answer pairs, the generation of $\mathbf{x}_{k+1}$ is based on logical reasoning with the context information in $\mathbf{x}_{1:k}$ according to its pretrained knowledge.

Our aim is to design a new generative RS that tightly couples LLMs with the recommendation task by introducing user/item ID tokens (and token embeddings), such that user/item semantics (e.g., users' interests in item) can be accurately modeled for effective and efficient recommendations, and the encoded knowledge and reasoning ability of pretrained LLMs can be fully utilized simultaneously.

\subsection{Extension of User/Item Tokens}

\begin{figure}
\centering
\includegraphics[width=\linewidth]{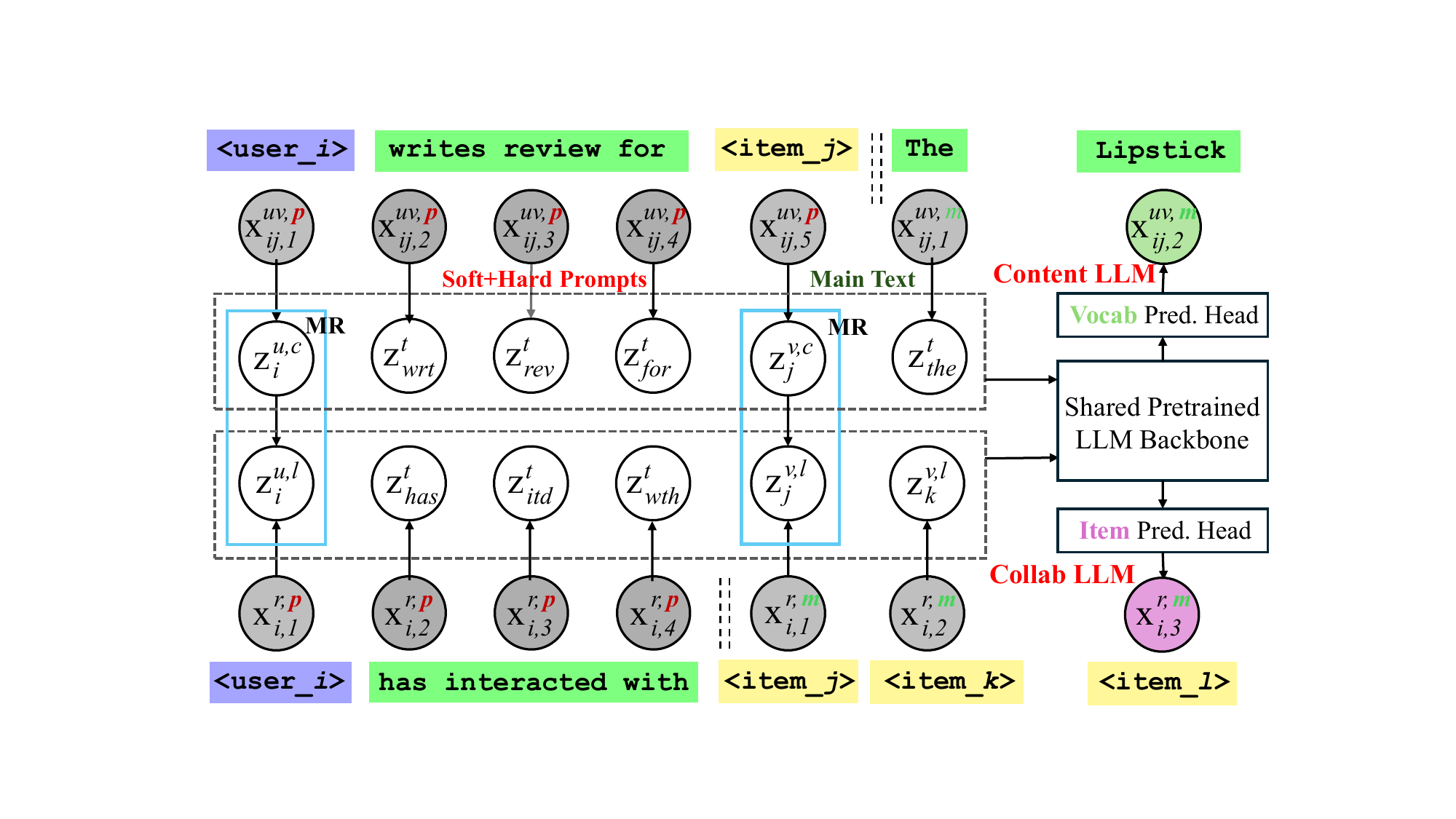}
\caption{The overview of the proposed CLLM4Rec in the mutually-regularized pretraining stage. Mutual regularization for item\_$\bm{k}$ is omitted for simplicity.}
\vspace{-3mm}
\label{fig:overall_pretrain} 
\end{figure}

\subsubsection{\textbf{Vocab Expansion}} To tightly couple the pretrained LLM with the recommendation task, we first expand the vocabulary of the LLM by adding user/item ID tokens to describe the intrinsic user/item semantics, such that the semantic gap between RS and natural language can be well bridged. We use bracket notations \textbf{"<user\_$\bm{i}$>"} and \textbf{"<item\_$\bm{j}$>"} to denote the newly-introduced token for the $i$-th user and the $j$-th item, which has token ID $N+i$ and $N+I+j$, and will not be broken down into atomic tokens.

\subsubsection{\textbf{Token Embeddings}}
\label{sec:token_emb}
For LLMs to understand the newly introduced user/item tokens, they must first be transformed into dense embeddings. Accordingly, we use $\mathbf{z}^{t}_{k} \in \mathbb{R}^{K}$ to represent the pretrained embedding of the $k$-th vocab token. In addition, for the newly introduced user/item tokens, we introduce \textit{two types of token embeddings} that are aligned with the vocab space to faithfully represent the user/item collaborative and content semantics. Specifically, we first sample user/item collaborative token embeddings from the same $K$-dimensional latent space as follows: 
\begin{equation}
\label{eq:ll_prior}
\mathbf{z}^{l, u}_{i}, \mathbf{z}^{l, v}_{j} \sim \mathcal{N} \left(\mathbf{0}, \lambda_{l}^{-1} \cdot \mathbf{I}_{K} \right), 
\end{equation}
where $\lambda_{l}$ is the prior precision for $\mathbf{z}^{l, u}_{i}, \mathbf{z}^{l, v}_{j}$. Importantly, to align the content semantics with the collaborative semantics for recommend-\\ation-oriented content modeling, we sample user/item content token embeddings from the following conditional prior: 
\begin{equation}
\label{eq:cc_prior}
\mathbf{z}^{c, u}_{i} \sim \mathcal{N}\left(\mathbf{z}^{l, u}_{i}, \lambda_{c}^{-1}\cdot \mathbf{I}_{K}\right), \mathbf{z}^{c, v}_{j} \sim \mathcal{N}\left(\mathbf{z}^{l,v}_{j}, \lambda_{c}^{-1}\cdot \mathbf{I}_{K}\right),
\end{equation} 
 where $\lambda_{c}$ is the precision for the conditional prior of $\mathbf{z}^{c, u}_{i}, \mathbf{z}^{c, v}_{j}$. The horizontally-stacked matrices of vocab/collaborative/content token embeddings are denoted as $\mathbf{Z}^{t}$, $\mathbf{Z}^{l, \{u,v\}}$, and $\mathbf{Z}^{c, \{u,v\}}$, respectively.

\subsubsection{\textbf{CLLM4Rec Base Model}}  With user/item tokens and the corresponding token embeddings introduced in the previous subsections, we are ready to introduce the CLLM4Rec base model with expanded vocabulary. The CLLM4Rec base model is denoted with 
\begin{equation}
\mathbf{h}^{(L)}_{\{l, c\}, 1:k} = \hat{llm}_{\{l, c\}}(\mathbf{x}_{1:k}),
\end{equation}
which maps the token sequence $\mathbf{x}_{1:k}$ into the hidden space $\mathbb{R}^{k \times K_{h}}$ through $L$ stacked self-attention modules (the superscript $(L)$ will be omitted if no ambiguity exists); here, $\mathbf{x}_{k}$ is a size $N+I+J$ one-hot vector denoting the token of either a vocab, a user, or an item. In addition, the subscript in $\hat{llm}_{\{l, c\}}$ denotes which embedding matrix is used to encode the user/item tokens (where $l$ stands for matrix $\mathbf{Z}^{l, \{u,v\}}$ and $c$ stands for matrix $\mathbf{Z}^{c, \{u,v\}}$). For the CLLM4Rec base model $\hat{llm}_{\{l, c\}}$, only the user/item token embeddings are trainable, whereas the vocab embeddings $\mathbf{Z}^{t}$ as well as the other parts of the backbone LLM are fixed to preserve the pretrained knowledge.

\subsection{Mutually-Regularized Pretraining}

With CLLM4Rec base model introduced in the previous section, we discuss the mutually-regularized pretraining strategy for CLLM4Rec. The aim is to learn user/item collaborative/content token embeddings based on language modeling on the corpora established from user-item interactions and user/item textual features, where the encoded knowledge and logical reasoning ability of the LLM can be fully utilized. The overall process can be referred to in Fig. \ref{fig:overall_pretrain}.

\subsubsection{\textbf{Recommendation-Specific Corpora}}
\begin{figure}
\centering
\includegraphics[width=0.83\linewidth]{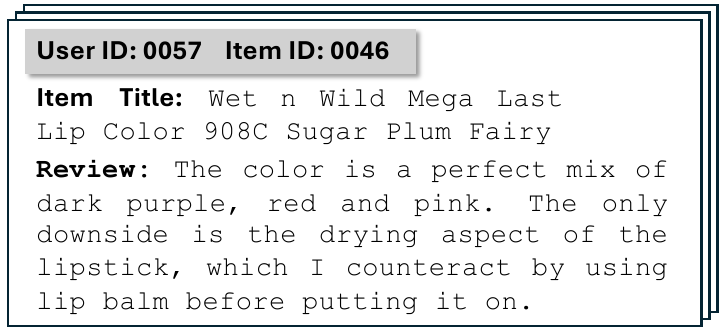}
\vspace{-3mm}
\caption{Exemplar review data from the Amazon Beauty dataset \cite{mcauley2016addressing}, where prior knowledge of natural language can help understand item property and user interests.}
\vspace{-2mm}
\label{fig:exp_review}
\end{figure}

Generally, we can transform the interactions $\mathbf{r}_{i}$ and user/item content features $\mathbf{x}^{u}_{i}$, $\mathbf{x}^{v}_{j}$,
$\mathbf{x}^{uv}_{ij}$ into documents of user/item/vocab token sequences as follows:

\vspace{2mm}
\begin{mdframed}[backgroundcolor=black!15] 
\textbf{Raw Corpora Transformed from Recommendation Data}

\vspace{1.5mm}
\noindent \textbf{(a) \textit{Historical Interactions} $\mathbf{r}_{i}$:} \\
\noindent \texttt{\small \colorbox{blue!35}{<user\_$i$>} \colorbox{green!50}{has interacted with} \colorbox{yellow!50}{<item\_$j$>  <item\_$k$> ...}}

\vspace{1mm}
\noindent \textbf{(b) \textit{User/Item Textual Features} $\mathbf{x}^{u}_{i}$, $\mathbf{x}^{v}_{j}$,
$\mathbf{x}^{uv}_{ij}$:}

\vspace{1mm}
\noindent \texttt{\small \colorbox{green!50}{The biography of}  \colorbox{blue!35}{<user\_$i$>} \colorbox{green!50}{is: Main biography.}} \\
\texttt{\small \colorbox{green!50}{The content of} \colorbox{yellow!50}{<item\_$j$>} \colorbox{green!50}{is: Main contents.}} \\ 
\texttt{\small  \colorbox{blue!35}{<user\_$i$>} \colorbox{green!50}{writes the review for} \colorbox{yellow!50}{<item\_$j$>}: \colorbox{green!50}{Main reviews.}} 
\end{mdframed}
\vspace{3mm}

\noindent where an example based on the Amazon Beauty dataset \cite{mcauley2016addressing} can be referred to in Fig. \ref{fig:exp_review}. However, directly conducting language modeling on the raw corpora is clearly infeasible, as each document is composed of heterogeneous vocab, user, and item tokens, where the number of meaningful vocab tokens (e.g., $\sim$ 50k for GPT, and $\sim$ 30k for T5) can be diluted by the large number of newly introduced user/item tokens with randomly initialized embeddings.

\subsubsection{\textbf{Soft$\bm{+}$Hard Prompting}}
\label{sec:soft+hard}
To address the above challenge, we propose a novel soft+hard prompting strategy to facilitate language modeling on RS-specific corpora with heterogeneous user/item/vocab tokens. The strategy is based on a key observation that documents transformed from both user-item interactions $\textbf{r}_{i}$ and user/item textual features $\mathbf{x}^{u}_{i}$, $\mathbf{x}^{v}_{j}$, $\mathbf{x}^{uv}_{ij}$ can be broken down into two parts: A \textit{heterogeneous} part composed of soft (user/item) and hard (vocab) tokens providing context information regarding the gist of the document, and a main text part with \textit{homogeneous} item/vocab tokens fulfilling the pretexts in detail. Therefore, we can view the first part as a soft+hard prompt and conduct language modeling only on the second part. This encourages the model to focus exclusively on collaborative and content information, such that the effectiveness and stability of language modeling can be substantially enhanced.

For collaborative modeling, document $\mathbf{x}^{r}_{i}$ transformed from the historical interactions of user $i$ can be broken down into the soft+hard prompt $\mathbf{x}^{r,p}_{i}$ and homogeneous item token sequence $\mathbf{x}^{r,m}_{i}$ as follows:

\vspace{2mm}
\begin{mdframed}[backgroundcolor=black!15] 
\noindent \textbf{(a) \textit{Historical Interactions} $\mathbf{r}_{i}$:} \\
\noindent $\underbrace{\texttt{\small \colorbox{blue!35}{<user\_$i$>} \colorbox{green!50}{has interacted with}}}_{\text{soft+hard prompt}  \  \mathbf{x}^{r, p}_i} \underbrace{\texttt{\small \ \colorbox{yellow!50}{<item\_$j$>  <item\_$k$> ...}}}_{\text{item token seq.} \ \mathbf{x}^{r, m}_i}$.\\
\end{mdframed}
\vspace{2mm}

\noindent Accordingly, we introduce the \textbf{collaborative LLM} by adding an item prediction head $f_{l} : \mathbb{R}^{K_{h}} \rightarrow \mathbb{P}(J)$ to the CLLM4Rec base model $\hat{llm}_{l}$, which maps the final-layer last-step hidden representation $\mathbf{h}_{l,-1}$ calculated via $\hat{llm}_{l}$ to the item probability space $\mathbb{P}(J)$ to predict the next item token. The weights of $f_{l}$ are tied with the item collaborative token embeddings $\mathbf{Z}^{l,v}$ as $f_{l}(\mathbf{h}_{l,-1}) = \operatorname{softmax}(\mathbf{Z}^{l,v} \cdot \mathbf{h}_{l,-1})$. The generative process of the collaborative LLM can be denoted as:
\begin{equation}
\label{eq:lm_clb}
\mathbf{x}^{r,m}_{i,k+1} \sim p^{f_{l}}_{\hat{llm}_{l}}\left(\mathbf{x}^{r,m}_{i,k+1}|\mathbf{x}^{r, m}_{i,1:k}, \mathbf{x}_{i}^{r, p}\right), 
\end{equation}
where the prompt $\mathbf{x}_{i}^{r, p}$ serves as a context to generate the next item token based on the previous item tokens. Since the generation of $\mathbf{x}^{r,m}_{i,k+1}$ requires attending to previous tokens, when maximizing the likelihood, the collaborative LLM pushes the token embeddings of user $i$, i.e., $\mathbf{z}^{l, u}_{i}$, and the token embeddings of the interacted items, i.e., $\mathbf{z}^{l, v}_{j}$, $\mathbf{z}^{l, v}_{k}$, $\cdots$, to be close to each other, where user/item collaborative semantics in recommendation can be accurately captured.

Similarly, for the document transformed from the user/item content $\mathbf{x}^{uv}_{ij}$, it can also naturally be split into a soft+hard prompt $\mathbf{x}^{uv,p}_{ij}$ and the main text $\mathbf{x}^{uv,m}_{ij}$ of homogeneous vocab token sequence as: 

\vspace{2mm}
\begin{mdframed}[backgroundcolor=black!15] 
\noindent \textbf{(b) \textit{User/Item Textual Features} $\mathbf{x}^{uv}_{ij}$:} \\
\noindent $\underbrace{\texttt{\small \colorbox{blue!35}{<user\_$i$>} \colorbox{green!50}{writes the review for} \colorbox{yellow!50}{<item\_$j$>}:}}_{\text{soft+hard prompt}\ \mathbf{x}^{uv, p}_{ij}}\underbrace{\texttt{\small \colorbox{green!50}{Main reviews.\vphantom{<item\_$j$>}}}}_{\text{vocab seq. $\mathbf{x}^{uv, m}_{ij}$}}$
\end{mdframed}
\vspace{2mm} 

\noindent Accordingly, we introduce the \textbf{content LLM} by adding a vocab prediction head $f_{c} : \mathbb{R}^{K_{h}} \rightarrow \mathbb{P}(N)$ to the CLLM4Rec base model $\hat{llm}_{c}$, which maps the final-layer last-step hidden representation $\mathbf{h}_{c, -1}$ calculated via $\hat{llm}_{c}$ (which shares the same pretrained LLM with $\hat{llm}_{l}$ but uses $\mathbf{Z}^{c, \{u, v\}}$ to decode the user/item tokens) to the vocab probability space. Similarly, the weights of $f_{c}$ are tied with the vocab embeddings $\mathbf{Z}^{t}$ as $f_{c}(\mathbf{h}_{c, -1}) = \operatorname{softmax}(\mathbf{Z}^{t} \cdot  \mathbf{h}_{c, -1})$. The generative process of the content LLM can be denoted as follows:
\begin{equation}
\label{eq:lm_cnt}
\mathbf{x}^{uv,m}_{ij,k+1} \sim p^{f_{c}}_{\hat{llm}_{c}}\left(\mathbf{x}^{uv, m}_{ij,k+1}|\mathbf{x}^{uv, m}_{ij,1:k}, \mathbf{x}_{ij}^{uv, p}\right),
\end{equation}
which generates the next vocab token $\mathbf{x}^{uv, m}_{ij,k+1}$ based on the previously generated vocab tokens $\mathbf{x}^{uv, m}_{ij,1:k}$ with prompt $\mathbf{x}_{ij}^{uv, p}$ as the context. When maximizing the likelihood, the content information in $\mathbf{x}_{ij}^{uv, m}$ can be encoded in the content token embeddings of user $i$ and item $j$, i.e., $\mathbf{z}^{c, u}_{i}$, $\mathbf{z}^{c, v}_{j}$, where the pretrained knowledge of the LLM can be fully utilized. For example, for the review shown in Fig. \ref{fig:exp_review}, the pretrained LLM will know that \textbf{<item\_$\bm{46}$>} is a lipstick with dark purple, red, and pink colors and can have side effects of drying lips, and reasons that \textbf{<user\_$\bm{57}$>} likes the colors but hates the side effects, which can be alleviated by applying lip balm. 

\vspace{2mm}

\noindent \textbf{Discussion.} Generally, since the "hard" (i.e., the vocab) part of the prompts $\mathbf{x}^{r, p}_{i}$ and $\mathbf{x}^{uv, p}_{ij}$ is what the pretrained LLM could understand, it is designed to trigger the reasoning ability of the pretrained LLM based on its encoded knowledge. For example, the relational phrase \textbf{"has interacted with"} in the prompt $\mathbf{x}^{r,p}_{i}$ guides the collaborative LLM to understand that the newly-introduced token \textbf{<user\_{$\bm{i}$}>} is a \textit{user subject} and the tokens in the prompt $\mathbf{x}^{r,m}_{i}$ are the \textit{objects} of interacted item sequences. Meanwhile, the contexts \textbf{"write the review for"} in $\mathbf{x}^{uv,p}_{ij}$ direct the content LLM to better understand the nature of main texts in $\mathbf{x}^{uv,m}_{ij}$, i.e., \textbf{<user\_$\bm{i}$>}'s judgment on \textbf{<item\_$\bm{j}$>} based on the personal using experience. The specific formulation of the prompt can be flexible, as Geng et al. \cite{geng2022recommendation} have demonstrated that variations in the expression of the prompt make less difference as long as the meaning is the same and the prompt is consistent across the training and testing phases. 

\subsubsection{\textbf{Mutually-Regularization}}

Since pretrained LLMs are not recommendation-oriented, naively optimizing Eq. (\ref{eq:lm_cnt}) unavoidably captures noisy information from content features irrelevant to recommendations. In addition, since user/item interactions are sparse, the collaborative LLM can easily overfit on the observed interactions when optimizing Eq. (\ref{eq:lm_clb}). To address these issues, we propose a mutually regularized pretraining strategy for CLLM4Rec, where collaborative LLM can guide content LLM to capture recommendation-related information from user/item content, and content LLM can in turn introduce side information to support collaborative filtering.

The mutual regularization naturally comes with the aligned generative process of CLLM4Rec defined in Eqs. (\ref{eq:ll_prior}), (\ref{eq:cc_prior}). Specifically, for user $i$, if we denote the stacked item token embeddings as $\mathbf{Z}^{c, v}_{i}$, $\mathbf{Z}^{l, v}_{i}$, which contains item $j$ and other items interacted by the user $i$, the generation process of CLLM4Rec associated with $\mathbf{x}^{r}_{i}$ and $\mathbf{x}^{uv}_{ij}$ can be defined as the joint distribution as follows:
\begin{equation}
\label{eq:joint}
\begin{aligned}
& p\left(\mathbf{x}^{r,m}_{i}, \mathbf{x}^{uv,m}_{ij}, \mathbf{z}^{l,u}_{i}, \mathbf{Z}^{l,v}_{i}, \mathbf{z}^{c, u}_{i},  \mathbf{Z}^{c, v}_{i} \big| \mathbf{x}^{r, p}_{i}, \mathbf{x}^{uv, p}_{ij}\right) = \\
& \underbrace{\Pi_{k} p^{f_{l}}_{\hat{llm}_{l}}\left(\mathbf{x}^{r, m}_{i,k} \big |\mathbf{x}^{r, m}_{i,1:k-1}, \mathbf{x}^{r, p}_{i}\right)}_{\text{\color{blue} \textbf{LM for collab. LLM}}} \cdot \underbrace{\Pi_{k} p^{f_{c}}_{\hat{llm}_{c}}\left(\mathbf{x}^{uv, m}_{ij,k} \big |\mathbf{x}^{uv, m}_{ij,1:k-1}, \mathbf{x}^{uv, p}_{ij}\right)}_{\text{\color[RGB]{0,128,0} \textbf{LM for content LLM}}} \cdot \\
& \underbrace{p\left(\mathbf{z}^{c,u}_{i}\big |\mathbf{z}^{l, u}_{i}\right) \cdot \Pi_{k} p\left(\mathbf{z}^{c,v}_{ik} \big | \mathbf{z}^{l, v}_{ik}\right)}_{\text{\color{purple} \textbf{mutual regularization}}} \cdot \underbrace{p\left(\mathbf{z}^{l, u}_{i}\right) \cdot \Pi_{k} p\left(\mathbf{z}^{l, v}_{ik}\right)}_{\text{\color{purple} \textbf{prior}}}.
\end{aligned}
\end{equation}
A scrutiny of Eq. (\ref{eq:joint}) reveals that the joint distribution can be decomposed into three parts: \textbf{\textit{(i)}} the language modeling of the collaborative and content LLMs that learn user/item token embeddings as Eqs. (\ref{eq:lm_clb}) and (\ref{eq:lm_cnt}); \textbf{\textit{(ii)}} the mutual regularization that connects the user/item token embeddings of the two LLMs (i.e., according to Eqs. (\ref{eq:ll_prior}), (\ref{eq:cc_prior}), $p\left(\mathbf{z}^{c,u}_{i}\big |\mathbf{z}^{l, u}_{i}\right)$ and $p\left(\mathbf{z}^{c,v}_{ik} \big | \mathbf{z}^{l, v}_{ik}\right)$ are conditional Gaussian, which will introduce MSE regularization between $\mathbf{z}^{c,u}_{i}, \mathbf{z}^{l, u}_{i}$, and $\mathbf{z}^{c,v}_{ik}, \mathbf{z}^{l, v}_{ik}$ when log-likelihood is maximized); \textbf{\textit{(iii)}} the prior of $\mathbf{z}^{l, u}_{i}$ and $\mathbf{z}^{l, v}_{ik}$, which will be ignored due to the existence of mutual regularization (i.e., setting the precision $\lambda_l$ in the prior in Eq. (\ref{eq:ll_prior}) as zero). 

We use Maximum a Posteriori (MAP) \cite{murphy2012machine} to estimate the user/item token embeddings $\mathbf{z}^{l,u}_{i}, \mathbf{Z}^{l,v}_{i}, \mathbf{z}^{c, u}_{i}, \mathbf{Z}^{c, v}_{i}$, where the objective is proportional to the logarithm of the joint distribution defined in Eq. (\ref{eq:joint}). Here, we take alternate steps to optimize the MAP objective. If we denote the trainable parameters associated with the item token prediction head $f_{l}$ and vocab token prediction head $f_{c}$ as $\boldsymbol{\theta}$ (which are tied with the corresponding token embeddings), the objective for the collaborative LLM (L-step) and content LLM (C-step) with mutual regularization can be derived as follows:

\vspace{2mm}
\noindent \textbf{L-step}. In the L-step, we fix user/item content embeddings $\mathbf{z}^{c, u}_{i}, \mathbf{Z}^{c,v}_{i}$ as $\hat{\mathbf{z}}^{c, u}_{i}, \hat{\mathbf{Z}}^{c,v}_{i}$ in Eq. (\ref{eq:joint}), and use them to constrain the user/item collaborative embeddings along with the language modeling of collaborative LLM, leading to the following composite objective:
\begin{equation}
\label{eq:l_step}
\begin{aligned}
&\mathcal{L}_{\mathrm{l\_step}}^{\mathrm{MAP}} \left(\mathbf{z}^{l,u}_{i}, \mathbf{Z}^{l,v}_{i}; \boldsymbol{\theta}\right)= \sum_{k} \underbrace{-\ln  p^{f_{l}}_{\hat{llm}_{l}}\left(\mathbf{x}^{r, m}_{i,k} \big |\mathbf{x}^{r, m}_{i,1:k-1}, \mathbf{x}^{r, p}_{i}\right)}_{\text{\color{blue} \textbf{LM loss for collab. LLM}}} \\
& \underbrace{+\frac{\lambda_c}{2} \left \|\mathbf{z}^{l,u}_{i} - \hat{\mathbf{z}}^{c, u}_{i} \right \|^{2}_{2} + \sum_{k} \frac{\lambda_c}{2} \cdot \left \|\mathbf{z}^{l,v}_{ik} - \hat{\mathbf{z}}^{c, v}_{ik}  \right \|^{2}_{2}}_{\text{\color{purple} \textbf{MR loss with content LLM}}}\  \underbrace{+\frac{\lambda_{l}}{2} \left\| \mathbf{z}^{l,u}_{i} \right \|^{2}_{2} + \frac{\lambda_{l}}{2} \left\| \mathbf{z}^{l,v}_{j} \right \| ^{2}_{2} \vphantom{\sum_{k} \frac{\lambda_b}{2}} }_{\text{\color{purple} \textbf{Prior loss}}} +\ \mathcal{C}_{l}, \\
\end{aligned}
\end{equation}
where $\mathcal{C}_{l}$ is the constant irrelevant for optimization. The \textbf{\textit{LM loss}} captures the collaborative similarity between token embeddings of user $i$ and the interacted items, where side information can be introduced via the \textbf{\textit{MR loss}} to support collaborative filtering.

\vspace{2mm}
\noindent \textbf{C-step}. After one-step optimization of the L-step, we fix the user/item collaborative token embeddings $\mathbf{z}^{l,u}_{i}$, $\mathbf{z}^{l,v}_{j}$ as $\hat{\mathbf{z}}^{l,u}_{i}$, $\hat{\mathbf{z}}^{l,v}_{j}$ in Eq. (\ref{eq:joint}), leading to the following composite objective for the content LLM:
\begin{equation}
\label{eq:c_step}
\begin{aligned}
\mathcal{L}_{\mathrm{c\_step}}^{\mathrm{MAP}}&\left(\mathbf{z}^{c, u}_{i}, \mathbf{z}^{c, v}_{j}; \boldsymbol{\theta}\right)=  \sum_{k} \underbrace{-\ln p^{f_{c}}_{\hat{llm}_{c}}\left(\mathbf{x}^{uv, m}_{ij,k} \big |\mathbf{x}^{uv, m}_{ij,1:k-1}, \mathbf{x}^{uv, p}_{ij}\right)}_{\textbf{\text{\color[RGB]{0,128,0} \textbf{LM loss for content LLM}}}} \\
& \underbrace{+\frac{\lambda_c}{2} \left \|\mathbf{z}^{c, u}_{i} - \hat{\mathbf{z}}^{l,u}_{i} \right \|^{2}_{2} + \frac{\lambda_c}{2} \cdot \left \|\mathbf{z}^{c, v}_{j} -\hat{\mathbf{z}}^{l,v}_{j} \right \|^{2}_{2}}_{\text{\color{purple} \textbf{MR loss with collab. LLM}}}\  + \ \mathcal{C}_{c}, \\
\end{aligned}
\end{equation}
where \textit{\textbf{MR loss}} encourages the content LLM to capture recommend-\\ation-oriented information from user/item textual features. In Eqs. (\ref{eq:l_step}) and (\ref{eq:c_step}), 
 $\lambda_c$ controls the strength of mutual regularization, which will be thoroughly discussed in the empirical study.

\subsubsection{\textbf{Stochastic Item Reordering}}

Another issue that hinders effective collaborative filtering via Eq. (\ref{eq:l_step}) is the order of item tokens when transforming the historical interactions $\mathbf{r}_{i}$ into a token sequence $\mathbf{x}^{r, m}_{i}$. Item order usually does not matter for direct recommendations as users' long-term interests can be viewed as fixed (even if it matters, the positional embeddings denoting the order of natural language may not capture the semantics of the order of interactions). To address this issue, we propose a stochastic item reordering strategy to randomly permute the item tokens in $\mathbf{x}^{r, m}_{i}$, with soft+hard prompt $\mathbf{x}^{r, p}_{i}$ fixed when optimizing the collaborative LLM as Eq. (\ref{eq:l_step}). Through this strategy, the order of items can be ignored without negative influence on the vocab tokens in $\mathbf{x}^{r, p}_{i}$.

\subsection{Recommendation-Oriented Finetuning}

\subsubsection{\textbf{Pretraining v.s. Finetuning}} The pretraining of CLLM4Rec aims to learn the user/item token embeddings based on the large corpora established from user-item interactions $\mathbf{r}_{i}$ and user/item textual features $\mathbf{x}^{u}_{i},\mathbf{x}^{v}_{j},\mathbf{x}^{uv}_{ij}$ via language modeling, such that prompts with heterogeneous user/item/vocab tokens can be properly understood by CLLM4Rec. However, for now, the pretrained CLLM4Rec can only complete item/vocab token sequences based on prompts, rather than making recommendations, and therefore the gap between NLP and RS is still not completely eliminated. In addition, naively treating the collaborative LLM as a recommendation model can lead to huge computational costs as the recommended items are sequentially generated via auto-regression. Therefore, we propose a novel recommendation-oriented finetuning strategy for CLLM4Rec, which aims to further finetune the pretrained collaborative LLM and tailor it for more efficient recommendations. 

\newcommand{\redcrossbox}{
  \begin{tikzpicture}[baseline=0.5ex]
    \draw[red] (0,0) -- (0.7,0.25);
    \draw[red] (0,0.25) -- (0.7,0);
    \draw[black] (0,0) rectangle (0.7,0.25);
  \end{tikzpicture}
}

\subsubsection{\textbf{Masked Prompting with Multinomial Prediction Head}}
\label{sec:finetune}
To achieve this purpose, we first design a masked prompting strategy to generate recommendation-oriented prompts and targets for CLLM4Rec finetuning. Specifically, for each user, we randomly mask the interacted items $\mathbf{r}_{i}$  by $100\times p_{m}\%$, where the remaining items are denoted as $\mathbf{r}^{masked}_{i}$. We then use $\mathbf{r}^{masked}_{i}$ to generate a recommendation-oriented prompt $\mathbf{x}^{rec, p}_{i}$ as the input. All  hold-out items, which we denote with a multi-hot vector $\mathbf{r}^{hold}_{i}$, are treated as the target. The prompt $\mathbf{x}^{rec, p}_{i}$ based on $\mathbf{r}^{masked}_{i}$ is designed as:
\vspace{2mm}
\begin{mdframed}[backgroundcolor=black!15] 
\noindent \textbf{(c) \textit{Masked Prompts \& Target for Finetuning}}\\
\texttt{\small (prompt) \colorbox{blue!35}{<user\_$i$>} \colorbox{green!50}{has interacted with} \colorbox{yellow!50}{<item\_$j^{\prime}$> \redcrossbox} \colorbox{yellow!50}{\redcrossbox <item\_$k^{\prime}$>} \colorbox{green!50}{the user will interact with:} \\(target) \colorbox{red!50}{$\mathbf{r}^{hold}_{i}\in \{0,1\}^{J}$} \hfill \redcrossbox (masked items)}
\end{mdframed}
\vspace{2mm}
which triggers the reasoning ability of the pretrained LLM by using relational phrase \textbf{"has interacted with"} to describe the historical interactions, and using the phrase \textbf{"the user will interact with"} to guide the prediction of the target hold-out items $\mathbf{r}^{hold}_{i}$. 

We name CLLM4Rec in the finetuning stage as \textbf{RecLLM}, which inherits the CLLM4Rec base model $\hat{llm}_{l}$ from the collaborative LLM in the pretraining stage and introduces a new item prediction head with multinomial likelihood, i.e., $f_{rec}$, whose weights are also tied with the item token embeddings $\mathbf{Z}^{l,v}$. The generation of the hold-out items $\mathbf{r}^{hold}_i$ via the RecLLM can be formulated as follows:
\begin{equation}
\label{eq:rec_gen}
\mathbf{r}^{hold}_i \sim multi\left(f_{rec}\left(\mathbf{h}^{rec}_{l,i,-1}\right), N^{hold}_{i}\right)\text{, where } \mathbf{h}^{rec}_{l,i} = \hat{llm}_{l}\left(\mathbf{x}^{rec, p}_{i}\right),
\end{equation}
where $multi$ denotes the multinomial distribution, and $N^{hold}_{i}$ is the number of hold-out items for user $i$. When finetuning the RecLLM according to Eq. (\ref{eq:rec_gen}), $\mathbf{h}^{rec}_{l,i,-1}$, which can be viewed as the latent variable summarizing the historical interaction of user $i$, is encouraged to be similar to the collaborative embeddings of all the interacted items. In addition, we keep it regularized with the content LLM in a similar manner as Eq. (\ref{eq:l_step})\footnote{The objective of the RecLLM is formulated in Eq. (\ref{eq:rec_step}) in Appendix \ref{sec:ft_obj}.}, and use the stochastic item reordering strategy to generate the prompt $\mathbf{x}^{rec, p}_{i}$. Through the proposed recommendation-oriented finetuning strategy, CLLM4Rec can efficiently generate recommendations in a single forward-propagation step while fully utilizing the encoded knowledge of the pretrained LLM backbone and the user/item token embeddings learned via mutually-regularized pretraining, where all $J$ items serve as the candidates. In addition, since the target $\mathbf{r}^{hold}_{i}$ is constrained to be in the item probability space, hallucinated items can be avoided.

\subsection{Predictions with CLLM4Rec}

After the pretraining and finetuning of CLLM4Rec, to make recommendations for user $i$, we can convert the \textit{whole} historical interactions of the user, i.e.,  $\mathbf{r}_{i}$, into the recommendation-oriented prompt $\hat{\mathbf{x}}^{rec, p}_{i}$ as described in Section \ref{sec:finetune} (with no masked items) and input it into the RecLLM model. Then, the multinomial probability $\hat{\mathbf{r}}_i$ \textbf{over all $J$ items} can be obtained through one forward propagation via $\hat{\mathbf{r}}_i = multi\left(f_{rec}\left(\hat{\mathbf{h}}^{rec}_{i,-1}\right)\right), \hat{\mathbf{h}}^{rec}_{i} = \hat{llm}_{l}\left(\hat{\mathbf{x}}^{rec, p}_{i}\right)$, where uninteracted items with top-$M$ scores in $\hat{\mathbf{r}}_i$ can be selected as recommendations.

\section{Empirical Study}

In this section, we present and analyze the experiments on four public datasets and the LinkedIn job recommendation dataset, aiming to answer the following three research questions:
\begin{itemize}[leftmargin=0.5cm]
    \item \textbf{RQ1.} How does CLLM4Rec, the first RS that tightly couples the ID-based paradigm with the LLM-based paradigm, perform compared to state-of-the-art ID-based and LLM-based RSs?
    \item \textbf{RQ2.} How does the \textbf{pretraining stage} of CLLM4Rec (including the mutual regularization trick and the stochastic item reorder strategy) influence the performance of CLLM4Rec?
    \item \textbf{RQ3.} How does the \textbf{finetuning stage} of CLLM4Rec with masked prompting and multinomial item prediction head influence the efficiency and effectiveness of recommendations?
\end{itemize}

\noindent Due to space limitation, we only discuss CLLM4Rec with GPT-2 \cite{radford2018improving} backbone in this section, which has 768-dimensional token embeddings and token size 50,257. Experiments with more LLM backbones are thoroughly discussed in Appendix \ref{sec:expt5}.

\subsection{Experimental Setup}

\subsubsection{\textbf{Datasets}} The four public datasets we include for experiments are Amazon (AM)-Beauty dataset, AM-Toys dataset, AM-Sports dataset \cite{mcauley2016addressing} and Yelp dataset \cite{zhou2020s3}. In preprocessing, we binarize the interactions by keeping only ratings > 3 and treat them as implicit feedback \cite{liang2018variational}. In addition, we filter the datasets such that they keep the 5-core property after binarization. For each user, we randomly select 80\% of interactions for training, 10\% for validation, and 10\% for testing, where at least one item is selected in the validation and the test set. The reviews users provide to the items are collected as the textual feature $\mathbf{x}^{uv}_{ij}$. The \textbf{real-world experiments} are based on a job recommendation dataset collected at LinkedIn, where users' clicks on the job Ads are logged as the implicit feedback, and users' self-provided biography $\mathbf{x}^{u}_{i}$ and the job descriptions $\mathbf{x}^{v}_{j}$ are collected as the textual features, respectively. The statistics of the dataset are summarized in Table \ref{tab:datasets} in the Appendix. 

\subsection{Comparison with Baselines}
\label{sec:baselines}

\subsubsection{\textbf{Baselines}} 
To demonstrate the multifaceted superiority of the proposed CLLM4Rec, we include the following ID-based and (L)LM-based RSs as baselines for comparisons:

\vspace{1mm}
\noindent \textbf{\textit{(i)} ID-based Baselines.}
\begin{itemize}[leftmargin=0.5cm]
    \item \textbf{\textsc{Multi-Vae}} \cite{liang2018variational} is an ID-based collaborative filtering baseline that recommends new items by reconstructing the ratings $\mathbf{r}_{i}$ via a variational auto-encoder (VAE) with multinomial likelihood.
    \item \textbf{\textsc{\textsc{Md-Cvae}}} \cite{zhu2022mutually} is a hybrid RS that extends the Multi-VAE by introducing a dual feature VAE on textual features to regularize the reconstruction of $\mathbf{r}_{i}$ in Multi-VAE.
\end{itemize}

\noindent \textbf{\textit{\textit{(ii)}} LM-based Baselines}\footnote{Note that both BERT4Rec and S$^{3}$Rec are original designed for sequential recommendation. In this paper, we use similar recommendation-oriented finetuning as CLLM4Rec to adapt them to direct recommendation, where item sequences generated from masked interactions are used to predict all hold-out items with multinomial likelihood.}.
\begin{itemize}[leftmargin=0.5cm]
\item {\textbf{\textsc{Bert4Rec}}} \cite{sun2019bert4rec} uses masked language modeling (MLM) proposed in BERT \cite{kenton2019bert} to learn user/item embeddings for recommendation via bidirectional self-attention. \item {\textbf{\textsc{S$^3$Rec}}} \cite{zhou2020s3} extends BERT4Rec by augmenting the MLM with auxiliary tasks such as item attribute prediction, where content features can be fused for self-supervised learning.
\end{itemize}

\noindent \textbf{\textit{\textit{(iii)}} LLM-based Baselines.} 

\begin{itemize}[leftmargin=0.3cm]
\item {\textbf{\textsc{Llm-Scratch}}} has the same structure as CLLM4Rec, but it trains the whole model from scratch instead of loading and fixing the weights of the pretrained LLM backbone. 
\item {\textbf{\textsc{Llm-CF}}} eliminates the content LLM from CLLM4Rec and the mutually-regularized pretraining step and uses only the collaborative LLM and RecLLM for recommendations.
\item {\textbf{\textsc{Llm-FtALL}}} has the same structure as CLLM4Rec, but it finetunes the whole network, including the vocab embeddings as well as other parts of the pretrained LLM, instead of training only the newly-introduced user/item token embeddings.
\item {\textbf{\textsc{Llm-FixOrd}}} has the same structure as CLLM4Rec, but it removes the stochastic item reordering strategy for both the collaborative LLM in pretraining and the RecLLM in finetuning.
\item {\textbf{\textsc{Llm-PreRec}}} discards finetuning and ranks the categorical probability from the next item token prediction head of the collaborative LLM in the pretraining stage to make recommendations.
\end{itemize}

\subsubsection{\textbf{Qualitative Analysis.}} 

For other existing LLM-based RSs (i.e., both pseudo-ID-based and description-based methods introduced in Section \ref{sec:llm4rec}), they represent users/items with multiple tokens and formulate direct recommendation as a next token generation problem. Since the generated tokens could be irrelevant to the recommendation purpose, candidate items usually need to be explicitly provided in the prompt to avoid hallucination (e.g., P5 \cite{geng2022recommendation} provides 100 candidate items where one is positive, and TALLRec \cite{bao2023tallrec} outputs yes/no decision based on user/item descriptions in the prompts, etc.). In contrast, CLLM4Rec can simultaneously generate multiple recommendations from the entire item candidate pool. Therefore, these methods cannot directly work in our setting, and the comparisons are mainly based on qualitative analysis.

\begin{table}[t]
\setlength{\tabcolsep}{2pt}
\vspace{-3mm}
\centering
\caption{Comparison between CLLM4Rec and various baselines with GPT-backbone on three Amazon Review datasets. }
\vspace{-2mm}
\label{tab:main_result}
\small
\begin{tabular}{lccc}
\toprule
{\textbf{AM-Beauty}} & \textbf{Recall@20} & \textbf{Recall@40} & \textbf{NDCG@100}   \\ \midrule
Multi-VAE  & 0.1295 & 0.1720 & 0.0835 \\ 
MD-CVAE    & 0.1472 & 0.2058 & 0.0976 \\ 
BERT4Rec   & 0.1126 & 0.1677 & 0.0781 \\
S$^{3}$Rec  & 0.1354 & 0.1789 & 0.0867 \\
\midrule
LLM-Scratch  & 0.0840 & 0.1265 & 0.0583 \\
LLM-CF       & 0.1319 & 0.1841 & 0.0855 \\
LLM-FtAll    & 0.1335 & 0.1988 & 0.0836 \\
LLM-FixOrd   & 0.1524 & 0.2219 & 0.1072 \\
LLM-PreRec   & 0.1547 & 0.2196 & 0.1051 \\
\midrule
CLLM4Rec      & \textbf{0.1656} & \textbf{0.2323} & \textbf{0.1118} \\ \bottomrule
\specialrule{0em}{-2pt}{-2pt}
\\
\toprule
{\textbf{AM-Toys}} & \textbf{Recall@20} & \textbf{Recall@40} & \textbf{NDCG@100}    \\ \midrule
Multi-VAE & 0.1076 & 0.1558 & 0.0781 \\ 
MD-CVAE   & 0.1291 & 0.1804 & 0.0844 \\ 
BERT4Rec  & 0.0853 & 0.1375 & 0.0532 \\
S$^{3}$Rec   &  0.1064 & 0.1524 & 0.0665 \\
\midrule
LLM-Scratch  & 0.0485 & 0.0771 & 0.0362 \\
LLM-CF       & 0.1027 & 0.1434 & 0.0680 \\
LLM-FtAll    & 0.1162 & 0.1542 & 0.0696 \\
LLM-FixOrd   & 0.1342 & 0.1887 & 0.0889 \\
LLM-PreRec   & 0.1308 & 0.1859 & 0.0874 \\
\midrule
CLLM4Rec      & \textbf{0.1436} & \textbf{0.1933} & \textbf{0.0918} \\ \bottomrule
\specialrule{0em}{-2pt}{-2pt}
 \\ \toprule 
{\textbf{AM-Sports}} & \textbf{Recall@20} & \textbf{Recall@40} & \textbf{NDCG@100}    \\ \midrule
Multi-VAE  & 0.0659 & 0.0975 & 0.0446 \\ 
MD-CVAE    & 0.0714 & 0.1180 & 0.0514 \\ 
BERT4Rec   & 0.0521 & 0.0701 & 0.0305 \\
S$^{3}$Rec & 0.0616 & 0.0813 & 0.0438 \\
\midrule
LLM-Scratch  & 0.0362 & 0.0538 & 0.0362 \\
LLM-CF       & 0.0642 & 0.0966 & 0.0419 \\
LLM-FtAll    & 0.0794 & 0.1002 & 0.0424 \\
LLM-FixOrd   & 0.0901 & 0.1295 & 0.0592\\
LLM-PreRec   & 0.0839 & 0.1248 & 0.0561 \\
\midrule
CLLM4Rec     & \textbf{0.0926} & \textbf{0.1351} & \textbf{0.0634} \\ \bottomrule
\end{tabular}
\vspace{-4mm}
\end{table}

\begin{table}[t]
\setlength{\tabcolsep}{2pt}

\centering
\caption{Comparison between CLLM4Rec and various baselines on the Yelp dataset and the LinkedIn dataset. }
\vspace{-5mm}
\label{tab:com_results}
\small
\begin{tabular}{lccc}
\\ \toprule 
{\textbf{Yelp}} & \textbf{Recall@20} & \textbf{Recall@40} & \textbf{NDCG@100}  \\ \midrule
Multi-VAE  & 0.0526 & 0.0842 & 0.0424 \\ 
MD-CVAE    & 0.0664 & 0.1058 & 0.0497 \\ 
BERT4Rec   & 0.0418 & 0.0724 & 0.0361 \\
S$^{3}$Rec     & 0.0563 & 0.0893 & 0.0485 \\
\midrule
LLM-Scratch   & 0.0199 & 0.0325 & 0.0159  \\
LLM-CF        & 0.0541 & 0.0860 & 0.0412 \\
LLM-FtAll     & 0.0653 & 0.0989 & 0.0520 \\
LLM-FixOrd    & 0.0694 & 0.1053 & 0.0524 \\
LLM-PreRec    & 0.0639 & 0.1021 & 0.0498 \\
\midrule
CLLM4Rec      & \textbf{0.0735} & \textbf{0.1149} & \textbf{0.0536} \\ \bottomrule

\specialrule{0em}{-2pt}{-2pt}\\
\toprule
{\textbf{LinkedIn}} & \textbf{Recall@10} & \textbf{Recall@20}  & \textbf{NDCG@10}    \\ \midrule
Two-Tower      & 0.1186 & 0.2041 & 0.0979 \\ 
\midrule
M6-Retrieval   & 0.1279 & 0.2118  & 0.1020       \\
CLLM4Rec-Emb   & 0.1302 & 0.2165 & 0.1034 \\
CLLM4Rec       & \textbf{0.1427} & \textbf{0.2398} & \textbf{0.1199} \\ \bottomrule
\end{tabular}
\vspace{-3mm}
\end{table}

\subsubsection{\textbf{Results on the Public Datasets}} 

We first analyze the experimental results on four public datasets to provide preliminary answers for \textbf{RQs.} \textbf{1, 2, 3}. From Tables \ref{tab:main_result} and \ref{tab:com_results}, we can find that the ID-base method, Multi-VAE, remains a strong baseline for collaborative filtering (CF). LLM-CF, the CF backbone of CLLM4Rec, cannot beat Multi-VAE on both AM-Sports and Toys datasets, even if the "hard" part of the prompt triggers the reasoning ability of the pretrained LLM. However, when utilizing the large textual data, CLLM4Rec outperforms its ID-based counterpart, MD-CVAE (which tightly couples an item content VAE with Multi-VAE) by a large margin. This is because MD-CVAE uses shallow bag-of-word representations of textual features, for which pretrained LLMs in CLLM4Rec can provide deeper understanding via their pretrained knowledge. The importance of pretrained knowledge can also be shown by the LLM-Scratch model, which performs the worst among all included baselines. An interesting finding is that, LLM-FtAll, which finetunes the whole model including the pretrained LLM backbone, performs worse than CLLM4Rec, which optimizes only the newly introduced user/item token embeddings. The reason could be that, since the weights of the pretrained LLM are fully optimized, the recommendation-specific corpora are still not enough to adapt the pretrained LLM with good generalization ability for RS. Therefore, the cons of degenerating the pretrained knowledge outweigh the pros of introducing extra RS-specific knowledge. We can also find that LLM-PreRec, which uses the collaborative LLM in the pretraining stage to generate recommendations, is already a strong baseline. This demonstrates the effectiveness of the soft+hard prompting strategy to facilitate efficient and stable language modeling on recommendation-oriented corpora with heterogeneous tokens. Still, CLLM4Rec performs better than LLM-PreRec, which demonstrates the effectiveness of recommendation-oriented finetuning in adapting collaborative LLM for efficient recommendations. 

\subsubsection{\textbf{Results on the LinkedIn Dataset}} In the real-world experiment, we compare CLLM4Rec with the two-tower (TT) model utilized in LinkedIn for job recommendations. The TT model is implemented as a two-branch multi-layer perceptron (MLP), where the input user/item embeddings include embeddings extracted from a graph neural network (GNN) learned on the user-job bipartite graph, as well as features extracted from an internal BERT model. In addition, since the textual features are available for almost every user and item, we compare CLLM4Rec with the state-of-the-art LLM-based RS, M6-Retrieval \cite{cui2022m6}, which takes the dimensional-reduced embeddings of user/item descriptions from M6 transformer for contrastive recommendations. The results are summarized in Table \ref{tab:com_results}. For Table \ref{tab:com_results}, we can find that CLLM4Rec outperforms the shallow TT model by a large margin. However, although the inference latency for CLLM4Rec is significantly improved compared with existing methods due to the introduction of recommendation-oriented finetuning, directly deploying CLLM4Rec online is still infeasible, as the inference budgets are higher compared to the TT model. Therefore, we design the CLLM4Rec-Emb baseline, which includes the user/item token embeddings $\mathbf{Z}^{l,u}$ and $\mathbf{Z}^{l, v}$ learned from CLLM4Rec (projected into 128 dimensions) as extra inputs for the TT model, which demonstrates a performance improvement than the original TT model and the M6-Retrieval model in our offline experiment. This demonstrates the potential application of CLLM4Rec in industrial applications where low latency matters.

\subsection{Parameter Sensitivity Analysis}
\label{sec:lambda}

\begin{figure}
  \centering
  \begin{subfigure}{0.45\linewidth}
    \centering
    \includegraphics[width=\linewidth]{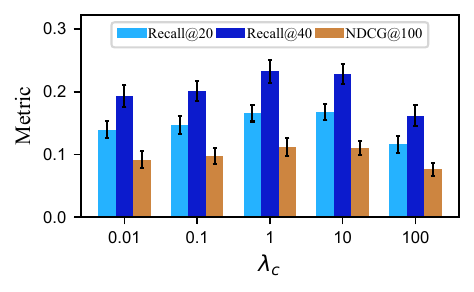}
    \vspace{-20pt}
    \caption{AM-Beauty Dataset}
    \vspace{2pt}
  \end{subfigure}
  \begin{subfigure}{0.45\linewidth}
    \centering
    \includegraphics[width=\linewidth]{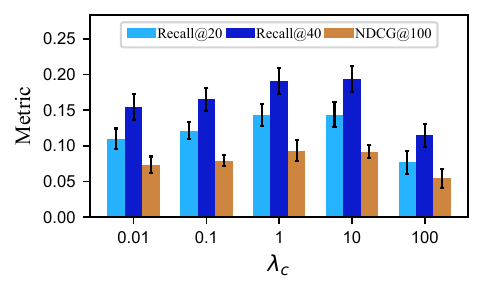}
    \vspace{-20pt}
    \caption{AM-Toys Dataset}
    \vspace{2pt}
  \end{subfigure}
  
  \begin{subfigure}{0.45\linewidth}
    \centering
    \includegraphics[width=\linewidth]{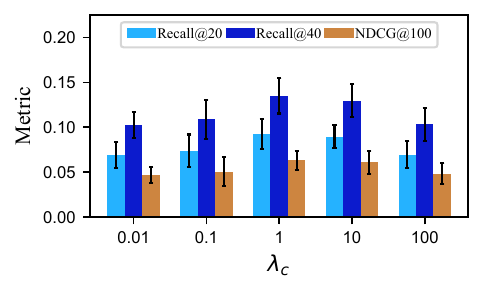}
    \vspace{-20pt}
    \caption{AM-Sports Dataset}
  \end{subfigure}
  \begin{subfigure}{0.45\linewidth}
    \centering
    \includegraphics[width=\linewidth]{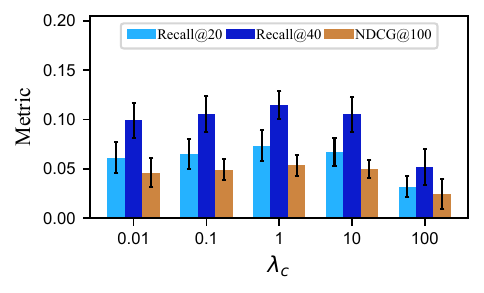}
    \vspace{-20pt}
    \caption{Yelp Dataset}
  \end{subfigure}
  \vspace{-2mm}
  \caption{Sensitivity analysis w.r.t. $\lambda_{c}$, which controls the strength of mutual-regularization for CLLM4Rec.}
  \label{fig:sens}
  \vspace{-5mm}
\end{figure}

To further answer \textbf{RQs. 2} and \textbf{3}, we vary $\lambda_{c}$ in Eqs. (\ref{eq:l_step}), (\ref{eq:c_step}), and (\ref{eq:rec_step}) that controls the strength of mutual regularization and investigates how it influences the performance of CLLM4Rec. From Fig. \ref{fig:sens}, we can find that, when $\lambda_{c}$ is small, the mutual regularization is weak, and the content LLM cannot provide enough user/item content side information to support the collaborative LLM and RecLLM. Therefore, the recommendation performance degenerates to a similar level as the LLM-CF. On the other hand, when $\lambda_{c}$ is too large, the MR loss in Eqs. (\ref{eq:l_step}), (\ref{eq:c_step}) and (\ref{eq:rec_step}) dominates, which hinders CLLM4Rec from learning useful user/item token embeddings via language modeling. Generally, for all four datasets, the performance of CLLM4Rec peaks at around $\lambda_{c}=1$, which serves as a good start when applying the GPT-based CLLM4Rec to new datasets.

\section{Conclusion}

In this paper, we proposed CLLM4Rec, the first method that tightly couples the ID paradigm and the LLM paradigm of RS, which faithfully captures user/item semantics while fully utilizing encoded knowledge and logical reasoning ability of pretrained LLMs simultaneously. Specifically, with mutually regularized pretraining based on soft+hard prompting strategy, CLLM4Rec can effectively capture the user/item collaborative and content information via language modeling. Furthermore, with recommendation-oriented finetuning, the pretrained knowledge of CLLM4Rec can be fully utilized to efficiently generate recommendations. Extensive experiments show the multifaceted superiority of CLLM4Rec over the state-of-the-art. 

\section*{Acknowledgment}
Yaochen Zhu and Jundong Li are supported in part by the National Science Foundation under grants (IIS-2006844, IIS-2144209, IIS-2223769, CNS2154962, and BCS-2228534), the Commonwealth Cyber Initiative Awards under grants (VV-1Q23-007, HV2Q23-003, and VV-1Q24-011), the JP Morgan Chase Faculty Research Award, and the Cisco Faculty Research Award.

\balance
\bibliographystyle{unsrt}
\bibliography{LLM4Rec}
\newpage
\appendix
\noindent \textbf{\Huge Appendix}

\begin{table}[h!]
  \caption{Statistics of the datasets. \#Feat. stands for number of textual features  (i.e., \# reviews for AM/Yelp datasets, and \#user biography+\#job descriptions for the LinkedIn dataset.}
  \label{tab:datasets}
 \vspace{-3mm}
  \begin{tabular}{lccccc}
    \toprule
    Dataset & \#Int. & \#Users & \#Items & Sparsity & \#Feat.\\
    \midrule
    AM-Beauty   & 94,148    & $10,553$  & $6,086$   & 99.85\%  & 70,604 \\
    AM-Toys     & 95,420    & $11,268$  & $7,309$   & 99.88\%  & 70,784 \\
    AM-Sports   & 185,718   & $22,686$  & $12,301$  & 99.93\%  & 137,618 \\
    Yelp        & 292,017   & $28,330$  & $18,775$  & 99.94\%  & 224,825 \\
    \midrule
    LinkedIn     & 90,173 & $22,391$  & $1,071$  & 99.62\% & 23,362\\
  \bottomrule
\end{tabular}
\end{table}

\section{Technical Details}
\subsection{Implementation of Soft+Hard Prompting}
 To implement the soft+hard prompting strategy discussed in Section \ref{sec:soft+hard}, for decoder-only LLMs such as GPT, we can generate only the "keys" and "values" for the heterogeneous tokens in the prompts $\mathbf{x}^{r, p}_{i}$, $\mathbf{x}^{uv, p}_{ij}$, and use the "query" of the last token as the start to generate the homogeneous tokens of the main texts $\mathbf{x}^{r, m}_{i}$, $\mathbf{x}^{uv, m}_{ij}$ for language modeling. For encoder-decoder-based LLMs such as T5, a natural thought is to input the prompts $\mathbf{x}^{r, p}_{i}$, $\mathbf{x}^{uv, p}_{ij}$ in the encoder, and use the decoder to generate the main texts $\mathbf{x}^{r, m}_{i}$, $\mathbf{x}^{uv, m}_{ij}$.
 
\subsection{Mutually Regularized Objective for Recommendation-Oriented Finetuning}
\label{sec:ft_obj}
If we denote the multinomial probability obtained from the RecLLM prediction head $f_{rec}$ as $\hat{\mathbf{r}}^{hold}_{i}$, and denote the stacked item collaborative token embeddings of items interacted by user $i$ as $\mathbf{Z}^{l,v}_{i}$, the \textbf{rec-step} objective of the recommendation-oriented finetuning (regularized with the content LLM) can be formulated as:
\begin{equation}
\label{eq:rec_step}
\begin{aligned}
\mathcal{L}_{\mathrm{rec\_step}}^{\mathrm{MAP}}& \left(\mathbf{z}^{l,u}_{i}, \mathbf{Z}^{l,v}_{i}; \boldsymbol{\theta}\right)= \underbrace{-\sum_{k} r^{hold}_{ik} 
\ln \hat{r}^{hold}_{ik} }_{\text{\color{blue} \textbf{Multinomial NLL Loss}}}   \underbrace{+\frac{\lambda_{l}}{2} \left\| \mathbf{z}^{l,u}_{i} \right \|^{2}_{2} + \sum_{k} \frac{\lambda_{l}}{2} \left\|  \mathbf{z}^{l,v}_{ik} \right \|^{2}_{2} \vphantom{\sum_{k} \frac{\lambda_b}{2}} }_{\text{\color{purple} \textbf{Prior loss}}} \\
& \underbrace{+\frac{\lambda_c}{2} \left \|\mathbf{z}^{l,u}_{i} - \hat{\mathbf{z}}^{c, u}_{i} \right \|^{2}_{2} + \sum_{k} \frac{\lambda_c}{2} \cdot \left \|\mathbf{z}^{l,v}_{ik} - \hat{\mathbf{z}}^{c, v}_{ik}  \right \|^{2}_{2}}_{\text{\color{purple} \textbf{MR loss with content LLM}}}\ +\ \mathcal{C}_{rec}, \\
\end{aligned}
\end{equation}
where NLL stands for negative log-likelihood, and $\mathcal{C}_{rec}$ is the constant irrelevant for the optimization purpose. From the form of the multinomial NLL loss we can find that, when finetuning the RecLLM according to Eq. (\ref{eq:rec_step}), the $\mathbf{h}^{rec}_{l,i,-1}$ output by the CLLM4Rec base model $\hat{llm}_{l}$, which can be viewed as the latent variable summarizing the historical interaction of user $i$, is encouraged to be similar to the collaborative embeddings of all the interacted items.

\section{Experiments}
\label{sec:expt5}

\subsection{Statistics of the Datasets}
\label{sec:imp}
The statistics of the public datasets and the LinkedIn recommendation dataset in the main paper are summarized in Table \ref{tab:datasets}.
\subsection*{}
\begin{table}[h!]
\setlength{\tabcolsep}{2pt}
\vspace{-3mm}
\centering
\caption{Comparison between CLLM4Rec with more backbones and more baselines on three Amazon Review datasets. }
\vspace{-2mm}
\label{tab:t5_result}
\small
\begin{tabular}{lccc}
\toprule
{\textbf{AM-Beauty}} & \textbf{Recall@20} & \textbf{Recall@40} & \textbf{NDCG@100}   \\ \midrule
Multi-VAE   & 0.1295 & 0.1720 & 0.0835 \\
EASE     & 0.1325 &	0.1757 & 0.0904 \\ 
BPR      & 0.1391 & 0.1803 & 0.0862 \\ 
MD-CVAE     & 0.1472 & 0.2058 & 0.0976 \\
\midrule
BERT4Rec    & 0.1126 & 0.1677 & 0.0781 \\
S$^{3}$Rec  & 0.1354 & 0.1789 & 0.0867 \\
\midrule
CLLM4Rec-T5   & 0.1538 & 0.2105 & 0.1052 \\
CLLM4Rec-LLaMA & 0.1614 & 0.2297 & 0.1103 \\
CLLM4Rec-GPT2      & \textbf{0.1656} & \textbf{0.2323} & \textbf{0.1118} \\ \bottomrule
\specialrule{0em}{-2pt}{-2pt}
\\
\toprule
{\textbf{AM-Toys}} & \textbf{Recall@20} & \textbf{Recall@40} & \textbf{NDCG@100}    \\ \midrule
Multi-VAE & 0.1076 & 0.1558 & 0.0781 \\ 
EASE      & 0.1082 & 0.1561 & 0.0787 \\ 
BPR       & 0.1124 & 0.1579 & 0.0824 \\ 
MD-CVAE   & 0.1291 & 0.1804 & 0.0844 \\ 
\midrule
BERT4Rec  & 0.0853 & 0.1375 & 0.0532 \\
S$^{3}$Rec   &  0.1064 & 0.1524 & 0.0665 \\
\midrule
CLLM4Rec-T5   & 0.1328 & 0.1840 & 0.0851 \\
CLLM4Rec-LLaMA & 0.1369 & 0.1877 & 0.0896\\
CLLM4Rec-GPT2      & \textbf{0.1436} & \textbf{0.1933} & \textbf{0.0918} \\ \bottomrule
\specialrule{0em}{-2pt}{-2pt}
 \\ \toprule 
{\textbf{AM-Sports}} & \textbf{Recall@20} & \textbf{Recall@40} & \textbf{NDCG@100}    \\ \midrule
Multi-VAE  & 0.0659 & 0.0975 & 0.0446 \\ 
EASE       & 0.0694 & 0.1038 & 0.0501 \\
BPR        & 0.0756 & 0.1057 & 0.0539 \\ 
MD-CVAE    & 0.0714 & 0.1180 & 0.0514 \\ 
\midrule
BERT4Rec   & 0.0521 & 0.0701 & 0.0305 \\
S$^{3}$Rec & 0.0616 & 0.0813 & 0.0438 \\
\midrule
CLLM4Rec-T5   & 0.0845 & 0.1226 & 0.0589 \\
CLLM4Rec-LLaMA & \textbf{0.0938} & \textbf{0.1369} & \textbf{0.0648} \\
CLLM4Rec-GPT2      & 0.0926 & 0.1351 & 0.0634 \\ \bottomrule
\end{tabular}
\end{table}
\subsection{\textbf{Implementation Details for the GPT-2-based CLLM4Rec}}

In this section, we introduce the implementation details for the GPT-2 based CLLM4Rec used in the main paper. During the training stage, we first optimize the content LLM as Eq. (\ref{eq:lm_cnt}) via language modeling for 10 epochs to warm up the user/item content token embeddings. Then, in the mutually regularized pretraining stage, we alternately train the collaborative and content LLMs as specified in Eqs. (\ref{eq:l_step}) and (\ref{eq:c_step}) for 100 epochs. Finally, we conduct the recommendation-oriented finetuning for 150 epochs, where the RecLLM is monitored with metrics Recall@20, Recall@40, and NDCG@100 calculated on the validation set as with \cite{liang2018variational}. RecLLM with the best performance is logged and evaluated on the test set as the final results. The prior precision $\lambda_{c}$ in Eqs. (\ref{eq:l_step}) and (\ref{eq:c_step}) is an important hyperparameter that controls the strength of mutual regularization. In the main paper, we first fix its value to the optimal one found by grid search and compare it with other baselines in Section \ref{sec:baselines}, and then we discuss its influence in Section \ref{sec:lambda}.

\subsection{Additional Results}
\subsubsection{\textbf{Implementation Details for More Backbones}}

In the appendix, we report the experiments of CLLM4Rec with two more LLM backbones to demonstrate the generalization ability of the proposed CLLM4Rec. The first backbone we consider is the T5-base model \cite{raffel2020exploring}, which is an encoder-decoder-based LLM with 32,128 vocab tokens (the last 28 tokens are empty), and each token is associated with a 768-dimensional vocab embedding. Another backbone is the LLaMA-7B model \cite{touvron2023llama}, which has the same number of tokens as the T5 model (as both use the sentence-piece tokenizer), and each token is associated with a 4,096-dimensional token embedding. For the non-symmetric LLM models where the weights of the LM prediction head are not tied with the vocab token embeddings, we randomly initialize the weights for the item prediction head for collaborative LLM (see Eq. (\ref{eq:lm_clb}) for details) and learn them together with the item collaborative token embeddings in both the pretraining and finetuning stages. Model training generally follows similar steps as the model with GPT-2 backbone described in Section \ref{sec:imp}, where we first warm up the content LLM as Eq. (\ref{eq:lm_cnt}) for ten epochs. Then, we conduct the mutually-regularized pretraining as Eqs. (\ref{eq:l_step}), (\ref{eq:c_step}) for 100 epochs, and continue for the recommendation-oriented finetuning as specified by Eq. (\ref{eq:rec_step}) for 150 more epochs.

\subsubsection{\textbf{More Baselines}}

In addition, we report the experiments of two more strong ID-based baselines, i.e., EASE \cite{steck2019embarrassingly} and BPR \cite{rendle2009bpr} in this Appendix. Specifically, EASE improves over the Multi-VAE by introducing a single-layer auto-encoder with constraints of no self-reconstruction, which shows better generalization ability due to its reduced variance and more suitable inductive bias faced with the sparse rating data. For the BPR model, we concatenate the user/item bag-of-word textual features with the user/item collaborative latent variables when optimizing the ranking-based objective. 

\subsubsection{\textbf{Results \& Analysis}}

The additional experimental results are summarized in Table \ref{tab:t5_result}. From Table \ref{tab:t5_result} we can find that, although CLLM4Rec-T5 generally outperforms the ID-based baselines and shallow LM-based baselines, its performance is consistently worse than the CLLM4Rec-GPT2 model. The reason for the overall inferior performance of CLLM4Rec with T5 backbone can be two-fold. First, we note that the weights in T5 are initialized with unit variance, whereas weights in GPT-2 are initialized with a variance of 0.02. Therefore, weights in T5 have much larger numerical values, which leads to large update steps. Therefore, the training of  CLLM4Rec-T5 is not as stable as CLLM4Rec-GPT2. In addition, in the finetuning stage of T5, the prompts are generally used to guide the macro behavior of the model. e.g., changing model behavior from question answering to machine generation via prompt "Translate English to French:". Therefore, another reason could be the mismatch between the original T5 prompts and the prompts intended to be used in CLLM4Rec. In addition, we can find that CLLM4Rec with the larger LLaMA-7B backbone cannot outperform CLLM4Rec-GPT2 on two smaller AM-Beauty and AM-Sports datasets, where the model can overfit on limited data. However, CLLM4Rec-LLaMA performs slightly better than CLLM4Rec-GPT2 on the comparatively large AM-Sports dataset, which demonstrates the generalization ability of CLLM4Rec with both larger models and larger data.

\end{document}